\documentclass[aps,prd,twocolumn,amsmath,amssymb,nofootinbib,preprintnumbers]{revtex4}

\usepackage[pdftex]{graphicx}
\usepackage{epstopdf}
\usepackage{amsmath}
\usepackage{amssymb}
\usepackage{subfigure}
\usepackage{hyperref}
\usepackage{url}
\usepackage{xcolor}
\usepackage{color}
\usepackage{graphicx}
\usepackage{gensymb}
\usepackage{tabularx}

\begin{document}
\title{\textbf{Pre-Hawking radiation cannot prevent the formation of apparent horizon}}

\author{
\textsc{Pisin Chen$^{1,2,3}$}\footnote{{\tt pisinchen{}@{}phys.ntu.edu.tw}},\;
\textsc{William G. Unruh$^{4}$}\footnote{{\tt unruh{}@{}physics.ubc.ca}},\;
\textsc{Chih-Hung Wu$^{1,2}$}\footnote{{\tt b02202007{}@{}ntu.edu.tw}}\;
and\;
\textsc{Dong-han Yeom$^{1,5,6}$}\footnote{{\tt innocent.yeom{}@{}gmail.com}}
}

\affiliation{~\\
$^{1}$\small{Leung Center for Cosmology and Particle Astrophysics, National Taiwan University, Taipei~10617, Taiwan}\\
$^{2}$\small{Department of Physics, National Taiwan University, Taipei 10617, Taiwan}\\
$^{3}$\small{Kavli Institute for Particle Astrophysics and Cosmology,
SLAC National Accelerator Laboratory, Stanford University, Stanford, California 94305, USA}\\
$^{4}$\small{Department of Physics and Astronomy, University of British Columbia, Vancouver, BC, Canada V6T 1Z1}\\
$^{5}$\small{Asia Pacific Center for Theoretical Physics, Pohang 37673, Republic of Korea}\\
$^{6}$\small{Department of Physics, Postech, Pohang 37673, Republic of Korea}\\
}

\begin{abstract}
As an attempt to solve the black hole information loss paradox, recently there has been the suggestion that, due to semi-classical effects, a pre-Hawking radiation must exist during the gravitational collapse of matter, which in turn prevents the apparent horizon from forming. Assuming the pre-Hawking radiation does exist, here we argue the opposite. First we note that the stress energy tensor near the horizon for the pre-Hawking radiation is far too small to do anything to the motion of a collapsing shell. Thus the shell will always cross the apparent horizon within a finite proper time. Moreovall, the amount of energy that can be radiated must be less than half of the total initial energy (if the particle starts at rest at infinity) before the shell becomes a null shell and cannot radiate any more without becoming tachyonic. We conclude that for any gravitational collapsing process within Einstein gravity and semi-classical quantum field theory, the formation of the apparent horizon is inevitable. Pre-Hawking radiation is therefore not a valid solution to the information paradox.
\end{abstract}

\maketitle

\newpage

%\tableofcontents

\section{Introduction}

The origin of the particles and in particular the origin of the energy, in
black hole evaporation has been a longstanding issue, which is still not
entirely settled. However in the 70's already, calculations of the regularized energy-momentum
tensor gave a clear picture of the energy flow due to the quantum particle
creation by the black hole \cite{Davies:1976ei}. In this picture, the conformal anomaly played a
crucial role \cite{Davies:1977}. The 1+1 dimensional black hole  gives the clearest picture.
For a massless scalar field in 1+1 dimensions, the classical energy momentum
tensor has a zero trace. Going to null coordinates $(u,v)$, the components of the
energy momentum tensor $T_{uu}$ and $T_{vv}$ are independently conserved so that
$T_{vv,u}=0$ and $T_{uu,v}=0$, which implies that the former is
constant along the $v=const$ surfaces, while the latter is also
constant along the $u=const$ surfaces. The flux of radiation, written,
for example, as $\mu l^\mu l^\nu$, where $l^\mu$ is a null vector, is
given by $T^{uu}$ and $T^{vv}$, which are related to the conserved
components by $T^{uu}= T_{vv}/g_{uv}^2$ and similarly for
$T^{vv}$. For the Schwartzschild metric where $g_{uv}=(1/2)
(1-2m/ r)$, the flux diverges as one nears the horizon.

However, the conformal anomaly, which makes the trace of the regularized
energy momentum tensor non-zero, upsets this picture, leading to the picture
that for large $r$, the Hawking radiation is a positive energy flux directed
along the $u=const$ rays, while near the horizon the energy flux is negative
and directed along the $v=const$ rays into the horizon, with no flux along the
$u=const$ rays near the horizon. Calculations in 3+1
dimensions also support this picture \cite{Candelas:1980}.

Despite this long standing picture, numerous authors regularly believe that the flux of
Hawking radiation is not originated from the conformal anomaly, but from the
matter which collapses to form the black hole. Since the null rays that reach
infinity intersect the infalling matter exponentially, the closer it is to the horizon,
the later those rays reach infinity. One therefore has an exponentially increasing
density of radiation near the horizon. If this were the picture, then that
density of energy could well drastically alter the picture of the horizon and
might well result in non-formation of the horizon.

Recently, another couple of papers have discussed what is called pre-Hawking
radiation \cite{Barcelo:2006uw}. It is not at all clear to us what this means, but from the
calculations it seems to mean that matter falling into the black hole decays,
emitting a positive energy flux toward infinity, a flux which near infinity
looks a lot like the Hawking flux.

While the standard picture given above is the best approximation we have to
the actual evaporation of a black hole, based as it is on the semiclassical
effect of quantum matter of the gravitational field, it is an interesting
question to ask-- what would happen if we allowed the infalling matter which
is eventually supposed to form a black hole, were to emit outgoing photons to
carry away at least a part of the energy of the infalling matter. Could one
arrange things so that this outgoing flux would completely deplete the energy
of the infalling matter so that no horizon ever formed? The two papers \cite{Kawai:2013mda} and
\cite{Baccetti:2016lsb} answer this by yes. This paper, on the other hand, will answer no. The
outgoing null radiation cannot carry away all of the energy of the infalling
matter. Some will always remain to form a black hole.

We emphasize that our performing this calculation does not imply that we
believe that this scenario represents in any way the effects of quantum
mechanics on the formation of a black hole. The regularized quantum energy,
even in the very early stages of collapse, does not give any support to the
idea that the energy of the outgoing radiation comes from the infalling
matter \cite{Arderucio-Costa:2017etb}. The infalling matter has an energy-momentum tensor that is conserved
independently of the quantum radiation which rides on top of the classical
spacetime. The conservation of the quantum energy comes from things like the
radiation of both positive and negative fluxes by the conformal anomaly.

This paper is organized as follows. In Sec.~\ref{sec:rad}, we analyze the 
radiation from a plane in a flat spacetime to demonstrate some fundamental 
properties that a radiating body must hold even without gravity. In 
Sec.~\ref{sec:radgrav}, we repeat the analysis for the case of a 
gravitational collapsing shell and confirm that the radiation must again be 
turned off at some point during the process. In Sec.~\ref{sec:max}, we 
investigate the maximum amount of radiation from the shell, if we 
impose the condition that the shell should always be timelike or null. 
Finally, in Sec.~\ref{sec:dis}, we briefly comment its implication on 
the black hole information loss paradox. Throughout this paper, 
we follow the convention of $c=G=\hbar=1$.

\section{\label{sec:rad}Null radiation from a plane shell in flat spacetime}

To investigate the backreaction to a radiating body, let us first consider a radiating plane in flat spacetime. We consider a sheet of pressureless matter with density $\sigma(u)$ moving along the path $x=X(u)$ in a flat plane-symmetric spacetime with the metric
\begin{eqnarray}
ds^2=du^2+2dudx.
\end{eqnarray}
 We further assume that this sheet of matter emits radiation of massless particles in the positive $x$ direction with a positive intensity $\lambda(u)$. Then the energy-momentum tensor is described by
\begin{eqnarray}
T^{uu} &=& \sigma \delta\left(x-X\right),\\
	%\end{eqnarray}
%\begin{eqnarray}
T^{ux}=T^{xu} &=& \sigma \dot{X} \delta\left(x-X\right),\\
%\end{eqnarray}
%\begin{eqnarray}
T^{xx} &=& \sigma \dot{X}^2 + \lambda(u) \Theta\left(x-X\right),
\end{eqnarray}
where $\delta(z)$ is the Dirac delta-function and $\Theta(z)$ the Heaviside step-function defined as $\Theta(z)=1$ for $z\geq0$ and $\Theta(z)=0$ for $z<0$.

Applying the local energy-momentum conservation law to the system, i.e., $T^{\mu\nu}_{,\nu}=0$, we obtain
\begin{eqnarray}
T^{uu}_{,u}+T^{ux}_{,x}=0 \;\;\;\; &\to& \;\;\;\; \dot{\sigma}=0, \label{eq:planar2}\\
%\end{eqnarray}
%\begin{eqnarray}
T^{xu}_{,u}+T^{xx}_{,x}=0 \;\;\;\; &\to& \;\;\;\; \ddot{X}=-\frac{\lambda}{\sigma} \label{eq:planar2}.
\end{eqnarray}

%Now we ask the question:
If $\lambda(u)$ is positive even when $\dot X=-1/2$, then $\dot X$ will
become less than $-1/2$, and the length-squared of the tangent vector to the
shell will go negative. The shell will become tachyonic. We believe tachyonic
matter is unphysical.

It is interesting to note that by using the coordinate $u$ to define the
trajectory of the shell, one can describe shells that are always timelike, null
or tachyonic. Had one used the proper length along the shell, the description
would have become singular as the shell became null, making the analysis more
difficult.

%If the radiating sheet moves with the speed of light, i.e., $\dot{X} = -1/2$, would the recoil effects of radiation cause the sheet to accelerate? The answer is, of course, trivial: such a null sheet must not accelerate. If it accelerates, then it would turn tachyonic, which is not acceptable. %Thus from Eq.~(\ref{eq:planar2}), we may deduce important insights. If a radiating body is null, then $\ddot{X}$ should be zero; otherwise the sheet will become tachyonic.
%Since $\sigma$ is a constant, the only choice to satisfy this condition is that $\lambda$ approaches zero, and hence the radiation from the sheet must be switched off by the time $\dot{X} = -1/2$.

\section{\label{sec:radgrav}Null radiation from a gravitational collapsing shell}

Now we consider a collapsing timelike shell that radiates energy in the out-going 
direction. After showing several simple calculations, we argue that it cannot 
continue to radiate indefinitely without becoming spacelike, i.e., 
tachyonic, which is unphysical.

The spacetime inside the shell is described by the flat Minkowski space in the
Eddington-Finkelstein-type coordinates:
\begin{eqnarray}
ds_{-}^2=-du^2-2dudr+r^2 d{\Omega}^2,
\end{eqnarray}
while the spacetime outside the shell is described by the outgoing Vaidya metric \cite{Lindquist:1965zz}
\begin{equation}
ds_{+}^2=-\left(1-\frac{2m(U)}{r}\right)dU^2-2dUdr+r^2 d{\Omega}^2,
\end{equation}
where $u$ and $U$ are outgoing null directions for inside and outside the shell, respectively, $m(U)$ is the mass function for outside the shell as a function of $U$, and $r$ is the areal radius for inside and outside the shell.

These two metrics should be continuous at the shell. Hence, one can express $ds_{+}^{2}$ in terms of the time variable $u$ after introducing a suitable redshift factor. The metric outside the shell becomes
\begin{eqnarray}
ds_{+}^2 &=& -\left(1-\frac{2m(u)}{r}\right)\left(\frac{dU}{du}\right)^2du^2 \nonumber \\
&& -2 \left(\frac{dU}{du}\right)dr du+r^2 d{\Omega}^2.
\end{eqnarray}
Note that the $u$ coordinate is finite and regular even across the horizon,
which is not true of the coordinate $U$. Using $u$ as our coordinate will thus
allow us to describe the behavior of the shell (and of the spacetime) even at the horizon.

The shell itself is defined by its path $r=R(u)$. Because of continuity of the angular part of the induced metric on the shell, this same equation must apply to both outside and inside the shell. Note that if the shell is a null ingoing shell, we must have $dR/du=-1/2$.

The condition that the induced metric on the shell be the same from either side of the shell requires
\begin{eqnarray}\label{eq:rad}
1 + 2 \frac{dR}{du} = U'^{2} \left( 1 - \frac{2m}{R} \right) + 2 U'\frac{dR}{du},
\end{eqnarray}
(where $U'={dU/du}$) or equivalently
\begin{eqnarray}\label{eq:cont}
\frac{dR}{du}= \frac{1- U'^2 \left(1-2m/R\right)}{2 \left( U'-1 \right)}.
\end{eqnarray}

In order to make the full metric continuous across the shell, we replace the radial coordinate $r$ by a new coordinate $z$, such that the shell is located at $z=0$:
\begin{eqnarray}
r=\begin{cases} R + \frac{z}{U'} &  (z > 0), \\  
    R + z & (z \leq 0). \end{cases}  
\end{eqnarray}
The metric of the spacetime thus becomes
\begin{widetext}
\begin{eqnarray}
ds^2 &=& -\left[ \left( \left(1-\frac{2m}{R+z/U'}\right)U'^2 + 2 U'R' - \frac{2zU''}{U'} \right) \Theta(z) + \left( 1 + 2 R' \right) \left( 1-\Theta(z) \right) \right] du^2 \nonumber \\
&& - 2dudz + \left( R+\frac{z}{U'} \Theta(z) + z \left( 1-\Theta(z) \right) \right)^2 d\Omega^2,
\end{eqnarray} 
\end{widetext}
for which all components of the shell are continuous across the shell. 

By assuming that the shell is composed of dust, we can solve the Einstein
equations for the motion of the shell. Since the Einstein tensor involves 
second derivatives of the metric, and since the metric is continuous across the shell, one expects
\begin{eqnarray}
G^{\mu\nu} = G_{\mathrm{bulk}}^{\mu\nu} + G_{\mathrm{shell}}^{\mu\nu} \delta(z). %+ G_{\mathrm{cont}}^{\mu\nu} \delta'(z),
\end{eqnarray}
%where $G_{\mathrm{cont}}^{\mu\nu}$ should be automatically vanished due to the continuity of the metric. Indeed, by plugging Eq.~(\ref{eq:cont}), one can show that all terms proportional to $\delta'(z)$ vanishes.
Only the  $uu, \theta\theta,$ and $\phi\phi$ coordinates of
$G^{\mu\nu}_{shell}$ can be non-zero  due to 
spherical symmetry. The dust conditions imply that the angular components of
the Einstein tensor must be zero. Calculating the Einstein tensor from the metric, we have
\begin{eqnarray}
G^{uu}_{\mathrm{shell}} &=& \frac{2\left( U'-1 \right)}{R U'},\\
G^{\theta\theta}_{\mathrm{shell}} &=& \frac{U'^2 m + U'' R^2 - R U'^2 + R U'}{R^4 U'}.
\end{eqnarray}
%The tangential stress is zero since we assume the shell is dust and thus Eq. (16) is zero.
Zero tangential stress implies that
\begin{eqnarray}\label{eq:ein}
U'' = -\frac{U' \left( U' m-R U' + R \right)}{R^2}.
\end{eqnarray}

Thus Eq.(\ref{eq:ein}) and Eq.(\ref{eq:rad}) give two first order equations for
the variables $U'$ and $R$. We can also write them in terms of a second order
equation for $R$ instead.
%Now we can calculate the dynamics of the shell. 
$R''$ can be obtained by differentiating both sides of Eq.(\ref{eq:cont}). One can simplify $R'$ and $U''$ by using Eqs.(\ref{eq:cont}) and (\ref{eq:ein}). Eventually this relation simplifies to
\begin{equation}\label{eq:Rpp}
R'' = \frac{U'}{2R(U'-1)} \left\{ 2m'U' - \left[1 -  \left( 1 - \frac{2m}{R} \right) U' \right]^2 \right\}.
\end{equation}
where $U'$ can be obtained in terms of $R$ and its derivative from eq
\ref{eq:rad}.
This implies that if $m' < 0$, then $R''$ is always negative.

Now we define
\begin{equation}
\rho \equiv R'(u)+\frac{1}{2} = - \frac{U'}{2(1 - U')} \left[1 - U' \left( 1 - \frac{2m}{R} \right) \right].
\end{equation}
This new variable is useful especially if the shell approaches the null direction; as $R'$ approaches $- 1/2$, $\rho$ goes to zero. By using this new variable, one can rephrase $R''$ as follows:
\begin{eqnarray}
{\rho'}=R''= \frac{m'U'^2}{R(U'-1)}-\frac{2(U'-1)}{RU'} \rho^2.
\end{eqnarray}
As the shell approaches a null shell, Eq.~(\ref{eq:cont}) implies that $U'$ approaches $(1-2m/R)^{-1}$. Hence, we obtain
\begin{eqnarray}
{\rho'}=\frac{m'}{2m-4m^2/R}-\frac{4m}{R^2} \rho^2
\end{eqnarray}
Therefore, if $\rho$ goes to zero while the shell continues to radiate ($m' < 0$), then $\rho$ will become negative, which implies that $R' < -1/2$ and the shell is tachyonic. It is therefore reasonable to conclude that a physical shell cannot emit unrestricted amount of energy; its radiation must be turned off at certain stage of the emission process (i.e., $m'$ should approach zero).

\begin{figure}
\begin{center}
\includegraphics[scale=0.7]{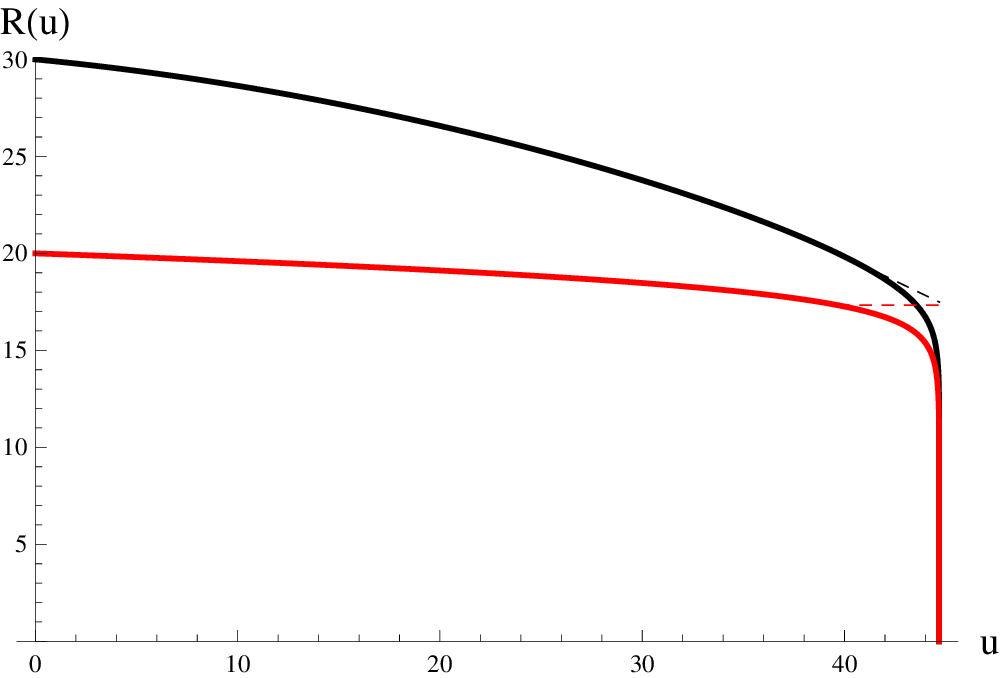}
\includegraphics[scale=0.75]{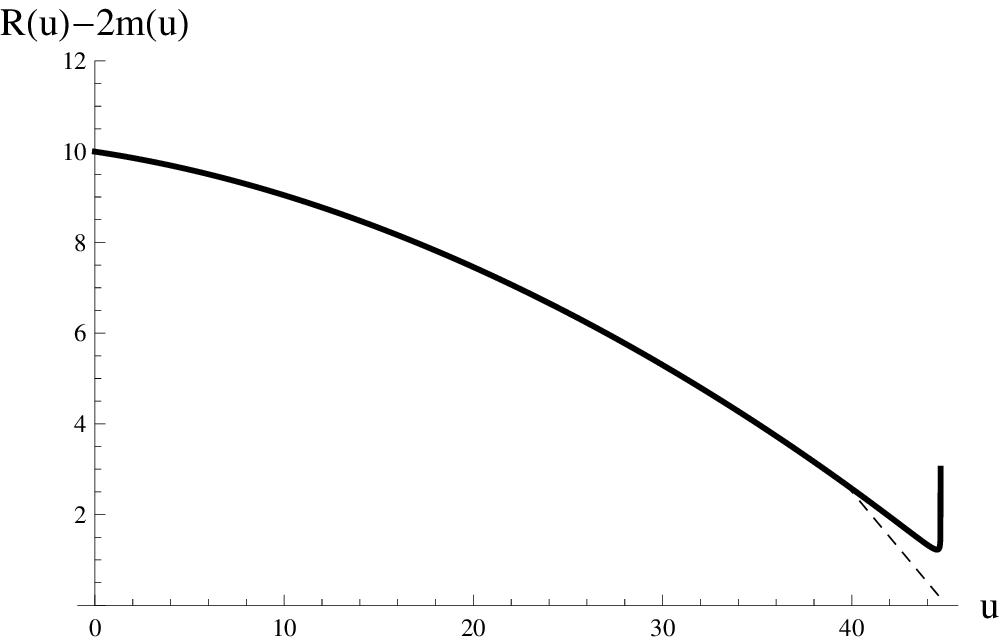}
\caption{\label{fig:R}Numerical evaluations for the coupled system of $R$,
$U$, and $m$, where the initial condition is given by $R(0) = 30$, $R'(0) =
-0.1$, $m(0) = 10$, $U(0) = 10$ is an arbitrary constant, and $U'(0)$ is given
by Eq.~(\ref{eq:cont}), where we choose the positive definite solution
for $U'$ since both $u$ and $U$ are future directed coordinates. The dashed lines are plots where $m'(u)$ is set to zero when the curve becomes lightlike so that the shell cannot go tachyonic. Upper: $R(u)$ (black) and $2 m(u)$ (red, apparent horizon). Lower: The solid lines shows that $R(u)-2m(u)$ is always positive and the shell does not cross the apparent horizon. The dashed curve does cross. From the upper curve the mass when the apparent horizon forms differs little from the initial mass.}
\end{center}
\end{figure}

\begin{figure}
\begin{center}
\includegraphics[scale=0.7]{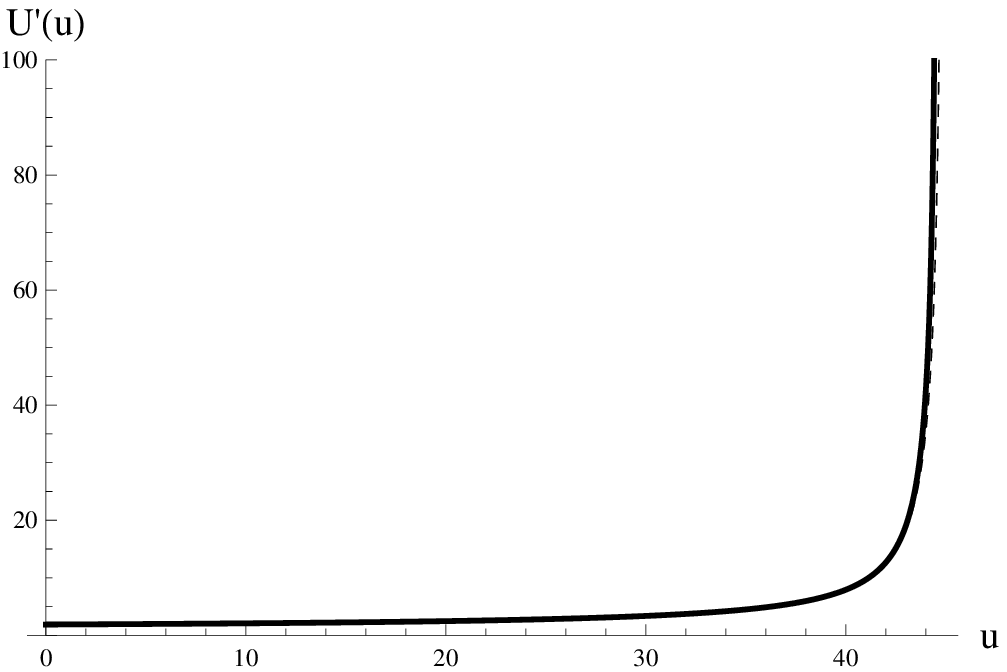}
\includegraphics[scale=0.7]{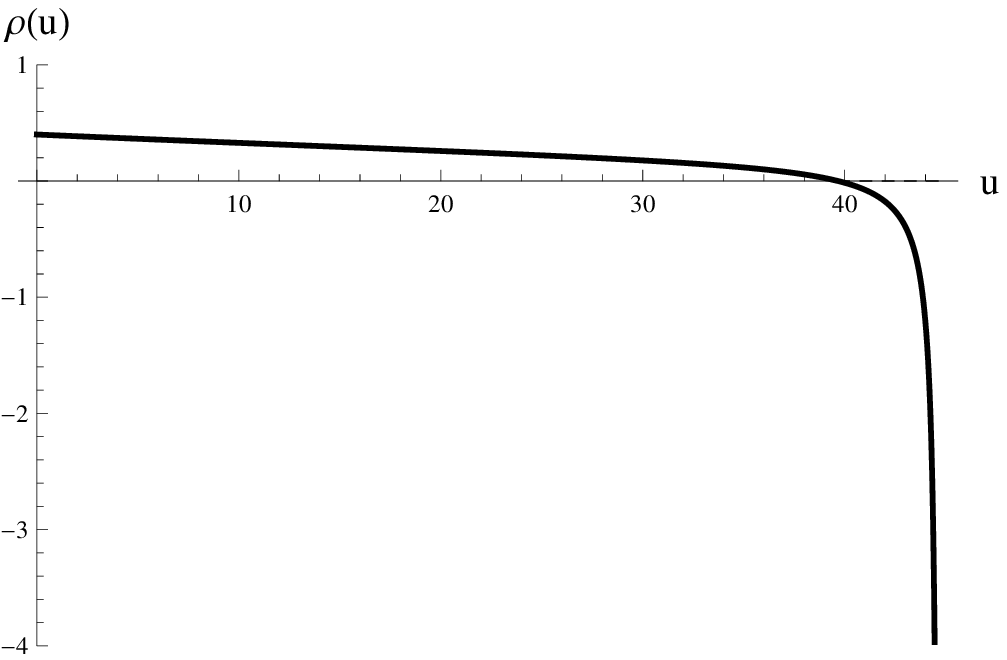}
\caption{\label{fig:rho}As the shell collapses, $U'(u)$ rapidly increases (upper). For the tachyonic case, $\rho(u)$ (lower) also goes large and negative, indicating a strongly tachyonic shell. For the dotted case, $\rho(u)$ is zero (lightlike shell) near the apparent horizon. Note that the dotted curve for $\rho(u)$ coincides with the $0$ axis.}
\end{center}
\end{figure}

Numerical evaluations for the coupled system of $R$, $U$, and $m$ as functions of $u$ are presented in Figs.~\ref{fig:R} and ~\ref{fig:rho}. One can solve Eqs.(\ref{eq:Rpp}) and (\ref{eq:ein}) for $R$ and $U$, respectively. 
In order to solve those equations we require a definite form for the function
$m(u)$. We follow the KYM and BMT models \cite{Kawai:2013mda,Baccetti:2016lsb} and assume that $m(u)$ satisfies,
\begin{eqnarray}
\frac{dm}{du} = - U' \frac{\alpha}{m^{2}},
\end{eqnarray}
where $\alpha$ is a numerical constant that depends on the number of fields that contribute to the Hawking radiation (we choose $\alpha = 1$ for numerical calculations). For initial conditions, we choose arbitrary constants for $R(0)$, $m(0)$, and $U(0)$, though $R(0) > 2m(0)$ so as to satisfy the condition that the shell is initially outside the apparent horizon. $R'(0)$ that represents the initial velocity of the shell is also arbitrary as long as $R'(0)>-1/2.$
Once a set of these initial conditions is chosen, $U'(0)$ is determined by Eq.(\ref{eq:cont}), where the positive solution is taken.

In our solution we do not assume that the shell remains timelike or null and allow it to become tachyonic.

Figs.~\ref{fig:R} and ~\ref{fig:rho} show the typical behavior of $R$ and $m$. As the shell approaches the apparent horizon, $U'$ increases rapidly and hence the radiation of the shell increases correspondingly. At first look, in Fig. 1 we see that a horizon never forms. The radius of the shell $R(u)$
always remains larger than $2m(u)$. However it is also clear from Fig. 1 that
$R'(u)$ becomes much smaller than $-{1/2}$, i.e., the shell becomes
tachyonic. If we set $m(u)'=0$ for $u$ larger than the point where $R'=-1/2$ (i.e., $m(u)$ thereafter remains constant), then $R'(u)$
remains equal to $-1/2$ and $R(u)$ will rapidly equal $2m(u)$ and continues to
$R(u)=0$, as shown by the dashed lines in the graphs.
.

\section{\label{sec:max}Maximum radiation by a particle}

One can get a feeling for the results of this section by looking at a massive
particle decaying to two oppositely directed photons. If the initial massive
particle is at rest, the two photons will each carry away half of the
rest-mass energy. If the massive particle is traveling in one direction, then the
photon emitted in that direction will have more than half of the energy, and
if the massive particle's velocity approaches that of light, then the energy in the
photon moving in the opposite direction will approach
zero. We will show that the same happens for the gravitating shell.

%Consider a particle with energy $E$ and velocity $v$ that emits a photon of energy $\epsilon$ in the opposite direction. The energy and velocity for the next moment are denoted by $E'$ and $v'$, respectively. Then
%\begin{eqnarray}
%E'=E-\epsilon,\\
%E'v'= Ev +\epsilon.
%\end{eqnarray}

%Requiring that the final-state of the particle be non-tachyonic, i.e., $v'\leq 1$, we have
%\begin{eqnarray}
%1\geq \frac{Ev+\epsilon}{E-\epsilon}= \frac{v+\epsilon/E}{1-\epsilon/E},
%\end{eqnarray}
%where the right hand side is an increasing function of $\epsilon/E$ and thus the maximum value of $\epsilon$ is $(1-v)/2$. Hence, the particle can radiate at most half of its initial energy in order not to become tachyonic.

Let us define the quantity
\begin{eqnarray}
%S &=& \left(\mathrm{det} g\right) g_{uu} \left( G_{\mathrm{shell}}^{uu} \right)^{2} \\
S \equiv 4 \frac{U'-1}{U'} R^2 \left[ 1 - U' \left(1-\frac{2m}{R} \right)\right].
\end{eqnarray}
This is the square of the energy density in the shell for timelike
matter and would be expected to be conserved if $m'=0$. It is however
also defined and conserved in the absense of radiation for null and
spacelike radiation as well.

Using the equations for $R'$ and $U'$, we find
\begin{eqnarray} \label{eq:Sprime}
S'= 8{m'} (U'-1) R,
\end{eqnarray}
which means $S$ is conserved if the radiation $m'$ is zero. %the mass remains unchanged.
If the shell is
null, $S=0$, and it is negative for tachyonic matter. 
%Since $S$ is defined in terms of $U$, $m$, and $R$, and not $g_{uu}$ explicitly, the condition $g_{uu}=0$ does not readily impose constraint on $S$. One may instead invoke $U'=(1-2m/R)^{-1}$ for a null shell to arrive at $S=0$.

We now evaluate $S$ under various situations. Let us first consider the
situation where a shell located at $R_0$  is at rest when it begins to collapse.
Since the shell satisfies $R' = 0$ at the starting point, from the equation for $R'$ we have
\begin{eqnarray}
0= 1-{U'_{0}}^{2} \left( 1- \frac{2m_{0}}{R_{0}} \right),
\end{eqnarray}
or
\begin{eqnarray}
U'_0= \frac{1}{\sqrt{1- 2m_0 / R_0}},
\end{eqnarray}
where the subscript $0$ denotes the initial condition of the shell. Then we obtain
\begin{equation}
S_{0} = 4R_{0}^{2} \left(\frac{U'_{0}-1}{U'_{0}}\right)^{2} = {16 m_0^2 \over\left(1+\sqrt{1 - 2m_0/R_0}\right)^2}.
\end{equation}
If $R_0 \gg 2m_0$, i.e., the shell falls in from a distance that is far from the putative horizon associated with $2m_0$, then the initial condition reduces to
\begin{eqnarray}
S_{0} = 4m_0^2.
\end{eqnarray}
If $S$ becomes negative, the shell turns spacelike, then it cannot represent physical matter. Thus if $S$ approaches zero, then $m'$ should reduce to zero correspondingly for the system to remain physical, as we argued earlier.

By integrating Eq.~(\ref{eq:Sprime}), and assuming that the minimum value of $S$ is $0$, we find
\begin{eqnarray}
-\Delta S \leq {16 m_0^2 \over\left(1 + \sqrt{1 - 2m_0/R_0}\right)^2}
\end{eqnarray}
and
\begin{equation}\label{eq:ineq}
{16m_0^2 \over\left(1+\sqrt{1 - 2m_0/R_0}\right)^2} \geq - 8\int (U'-1)R {dm\over du} du,
\end{equation}
where $\Delta S = \int S' du$ for a given integration domain and $m'$ is assumed to be negative. Since
\begin{eqnarray}
\frac{d (U'-1)R}{du}=\frac{(U'-1)^2}{2} \geq 0,
\end{eqnarray}
the multiplier of $m'$ is increasing and reaches its minimum at $R_0$. Therefore, $- \int (U' - 1) R dm \geq - (U'_{0} - 1) R_{0} \int dm$ and we have
\begin{eqnarray}
-\Delta m \leq \frac{m_0}{2} \frac{2\sqrt{1 - 2m_0/R_0}}{1+\sqrt{1 - 2m_0/R_0}} \leq \frac{m_0}{2}.
\end{eqnarray}
%which means the maximum amount of mass that can be radiated, before the shell becomes null and cannot radiate anymore, is less than half of the gravitational mass of the shell. Thus we have that $m(u)>m_0/2$ always and $R$ must go less than $2m$. The maximum radiated mass occurs if the radiation all comes out very near $R_0$; the closer the shell comes to $2m$, the less of the mass can be radiated without the shell going null. The maximum mass radiated is achieved if the initial radius goes to infinity, and the radiation occurs when the shell is near infinity. Therefore, one cannot prevent the black hole from forming by radiating away the mass of the shell in null radiation.
Note that the closer the shell approaches the horizon, the smaller the $\Delta m$ becomes. Also, from Eq.~(\ref{eq:ineq}), the closer to the horizon where the radiation is emitted, the smaller the change in mass must be. Thus, the
maximum amount of mass that can be radiated must be less than $1/2$ of the
total, and it must take place far from the horizon (where quantum effects would predict a negligible amount of radiation). Evidently, the shell cannot radiate away its entire
gravitational mass to prevent the formation of the apparent horizon.

Would a nonzero initial inward velocity at infinity of the shell help to ameliorate the situation? The answer is no. Using the equation for $R'$ again, we have
\begin{eqnarray}
R'_0 = \frac{1-{U'_{0}}^2(1-2m_0/R_0)}{2(U'_0-1)}.
\end{eqnarray}
Assuming $m_{0} \ll R_{0}$, we expand $U'_0$ as $U'_0 = 1 +\alpha m_0/R_0 + \mathcal{O} (m_{0}/R_{0})^{2}$ and $m_{0} \ll R_{0}$, and find
\begin{eqnarray}
\alpha = \frac{1}{1+R'_0}
\end{eqnarray}
and
\begin{eqnarray}
S_0 = 4m_{0}^{2} \alpha (2-\alpha) = 4m_0^2 \left( \frac{1+2R'_0}{1+R'_0}\right).
\end{eqnarray}
Hence, we obtain the following inequality
\begin{equation}
-8\int (U'-1)R m' du \leq - \left(\frac{8m_0}{1+R'_0}\right) \Delta m.
\end{equation}
Thus we have
\begin{eqnarray}
-\Delta m \leq \left(\frac{1+2R'_0}{2}\right) m_0 \leq \frac{m_0}{2},
\end{eqnarray}
since $-1/2 \leq R'_0 \leq 0$. Again, the maximum amount of radiation that can be radiated is half of the initial Schwarzschild mass, which corresponds to the situation where $R'_0=0$ at infinity.

If the shell becomes null, it has no rest mass. Thus the radiation can cause the rest mass of the shell to go to zero. However, it is the energy of the shell, not its rest mass, that determines its Schwarzschild mass, and the gravitational mass of the shell cannot go to zero without the shell going tachyonic.

\section{\label{sec:dis}Conclusion}

In this paper, we critically examined the assertion that a radiating and collapsing shell can never cross the apparent horizon. Some authors refer to such a radiation as pre-Hawing radiation. Our analysis has resulted in two major conclusions. First, the collapsing shell may emit radiation before the horizon is formed, but it must be weak unless the shell becomes spacelike. Second, the maximum amount of the radiated energy is bounded, where our estimation shows that it cannot be larger than half of its initial energy. As time goes, either the radiation stops or the shell becomes tachyonic. Since the latter is unphysical, it is inevitable that the radiation stops at some point. 

Although our analysis does not rely on the quantum mechanical nature of the pre-Hawking radiation, authors of the recent work \cite{Arderucio-Costa:2017etb} indeed show that quantum stress-energy tensor cannot play an important role during gravitational collapse, which is consistent with our result.
In a different setup, authors of \cite{Padmanabhan:2009} 
calculate a 2-D model of the quantum radiation from a massless scalar field for a
collapsing shell that stops (due to transverse stresses) just outside the horizon. They show that the quantum emission is not sufficient to stop the system from almost having a horizon very near the expected value associated with the mass that began the collapse. Their conclusion is consistent with ours.

We conclude that the assertion that pre-Hawking radiation can prevent the formation of the apparent horizon is based on unphysical assumptions. Black holes do form. 

%There can be one exception, however. Roughly speaking, it takes approximately $\sim M$ order of time for a black hole to form singularity, while it takes $\sim M^{3}/N$ order of time to evaporate, where $N$ is the number of fields that contribute to Hawking radiation. If these two time scales are of the same order, i.e., if $M \lesssim \sqrt{N}$, then the collapsing matter may indeed not form a black hole due to the very strong pre-Hawking radiation. We see that both of KMY and BMT models would be irrelevant unless the collapsing system is around the Planck scale. On the other hand, it would be invalid to apply semi-classical arguments about black holes at the Planck scale.

\section*{Acknowledgment}

PC is supported by ROC Ministry of Science and Technology (MOST), National Center for Theoretical Sciences (NCTS), and Leung Center for Cosmology and Particle Astrophysics (LeCosPA) of National Taiwan University. PC is in addition supported by U.S. Department of Energy under Contract No. DE-AC03-76SF00515. WGU thanks the Leung Center for Cosmology for their
hospitality and support  where his involvement in this
project began, Perimeter Institute (supported by the taxpayers of Ontario and
Canada amongst others) for their hospitality and support during some of the
work,  the Canadian Institute for Advanced Research for
their support, and the Natural  Science and Engineering Research Council of Canada
for their research support in the form of an operating grant.
DY was supported by Leung Center for Cosmology and Particle Astrophysics (LeCosPA) of National Taiwan University (103R4000).

%\newpage

\end{document}